\documentstyle[12pt]{article}  
\newcommand{\be}{\begin{equation}} 
\newcommand{\ee}{\end{equation}} 
\newcommand{\bea}{\begin{eqnarray}} 
\newcommand{\eea}{\end{eqnarray}}

\newcommand{\RR}{\rangle} 
\newcommand{\LL}{\langle}

\makeatletter \newdimen\normalarrayskip  
\newdimen\minarrayskip  
\normalarrayskip\baselineskip \minarrayskip\jot \newif\ifold             
\oldtrue           
  
\makeatother  
\newlength{\extraspace} 
\newlength{\extraspaces} 
\begin{document}  
\addtolength{\baselineskip}{.8mm}   
\thispagestyle{empty}  
\begin{flushright} 
\baselineskip=12pt  gr-qc/0009083\\ 
\hfill{  } 
\\September 2000 
\end{flushright} 
\vspace{.5cm}  
\begin{center}  
\baselineskip=24pt  
{\Large\bf{Information Erasure and 
\\the Generalized Second Law of Black Hole Thermodynamics}}\\[15mm]  
\baselineskip=12pt  
{\bf David D. Song} \\[3mm] 
{\it Centre for Quantum Computation, Clarendon Laboratory\\ 
University of Oxford, Parks Road, Oxford OX1 3PU, U.K.\\
{\tt d.song@qubit.org}} \\[6mm] 
{\bf Elizabeth Winstanley}\\[3mm] 
{\it Department of Applied Mathematics, University of Sheffield,\\ 
Hicks Building,  Hounsfield Road, Sheffield S3 7RH, U.K.\\
{\tt E.Winstanley@sheffield.ac.uk}}\\[40mm]   
{\sc Abstract}  
\begin{center} 
\begin{minipage}{15cm}  
We consider the generalized second law of black hole 
thermodynamics in the light of quantum information theory,
in particular information erasure and Landauer's principle
(namely, that erasure of information produces at least
the equivalent amount of entropy).
A small quantum system outside a black hole
in the Hartle-Hawking state is studied, and the quantum system
comes into thermal equilibrium with the radiation surrounding
the black hole.
For this scenario, we present a simple
proof of the generalized second law based on quantum
relative entropy.
We then analyze the corresponding information erasure
process, and confirm our proof of the generalized second law
by applying Landauer's principle.
\end{minipage} 
\end{center}  
\end{center}  
\noindent \vfill \newpage 
\pagestyle{plain} 
\setcounter{page}{1}  
The correspondence between the laws of thermodynamics 
and black hole mechanics was noted, 
as a curiosity without physical implications,
in a seminal paper by Bardeen, Carter and Hawking \cite{bch}.
At around the same time, Bekenstein \cite{bekenstein1}
was advocating a rather more radical approach. 
Noting the area theorem of black holes, which states that the
total area of black hole event horizons can never decrease, 
he observed that this is analogous 
to the ordinary second law of thermodynamics, 
i.e.  the total entropy of a closed system never decreases.  
He proposed that, multiplied by appropriate powers of the Planck length,  
Boltzmann constant and some dimensionless constant of order unity, 
the black hole area 
should be interpreted as its physical entropy.   
This proposal was given physical support by the discovery of
Hawking \cite{hawking} that black holes 
radiate at a temperature 
\be 
T_{bh} = \frac{\kappa}{2\pi} 
\label{temperature}
\ee 
where $\kappa$ is the surface gravity (here, and throughout the paper,
we work in Planck units, in which $\hbar = c=G=k_{B}=1$, where
$k_{B}$ is Boltzmann's constant).
This, coupled with the first law of black hole mechanics \cite{bch},
gives the value of the numerical constant  
in Bekenstein's  conjecture of black hole 
entropy to be $\frac{1}{4}$.   

Wheeler provided the initial motivation 
for Bekenstein's black hole entropy proposal \cite{bekenstein2}.  
Wheeler suggested a creature, subsequently called  Wheeler's demon, 
which could violate the ordinary second law of thermodynamics by 
dropping entropy into a black hole, producing a decrease in
the entropy outside the black hole.  
This led Bekenstein to conjecture that the black hole itself has 
an entropy (proportional to the area of the event horizon)
and, furthermore, the sum of the entropy outside the black hole and 
the black hole entropy must not decrease, 
\be 
\Delta S_{out} + \Delta S_{bh} \geq 0. 
\label{S_out}
\ee 
This generalized second law has been widely discussed in the literature,
and there are proofs due to Frolov and Page \cite{page}, 
and, more recently Mukohyama \cite{muko}.
Both these proofs make use of quantum field theory in curved space, and 
apply to quasistationary black holes.  
Frolov and Page's proof is applicable to eternal black hole space-times,
whilst Mukohyama considers black holes arising from gravitational
collapse.

In this paper we wish to consider the generalized second law
from another point of view, namely quantum information
theory and, in particular, Landauer's principle of
information erasure \cite{landauer}.
We will consider a quantum system outside a black hole,
which then comes into thermal equilibrium with the Hawking
radiation surrounding the black hole.
This scenario is different from those that
have been considered previously in proofs of the
generalized second law, giving further weight
to its validity.
 
We firstly discuss Maxwell's demon \cite{maxwell}, 
which is the analogue in ordinary thermodynamics of Wheeler's demon.  
Consider a container of gas which is divided into
two halves, left and right, by a partition.
Imagine now that there is a demon sitting on the partition, 
who is able to measure the 
velocities of individual molecules in the gas. 
If the demon let 
the fast molecules move to the  right container while 
keeping the slower ones to the left,  
then this would create a temperature difference 
and violate the second law of 
thermodynamics. 
Bennett noted \cite{bennett} that in order to do free work, the 
demon has to record its measurement result, and then its memory 
needs to be erased in order to do the next measurement.   
Landauer's principle states that in order to erase a certain amount of 
information at least the same amount of entropy must be 
generated. 
Therefore, the erasure of the memory of the demon 
generates an entropy greater than or equal to
the amount of recorded memory, which preserves the second law. 
Bennett's classical analysis of Maxwell's demon was 
later confirmed quantum mechanically \cite{zurek,lubkin,lloyd}. 
This process resembles Wheeler's demon, who is trying to erase 
information by dropping an object into a black hole.
This necessarily creates an increase of black hole entropy by 
at least the same amount as the dropped entropy, according to the
generalized second law. 
In this paper, we give a simple proof of the generalized 
second law for our model, 
using known results on quantum relative entropy. 
This will be confirmed by our analysis of the
corresponding information erasure process using Landauer's principle.
The generalized second law has recently been considered in the context
of quantum information theory (concentrating on the entanglement
of states inside and outside the black hole event horizon) 
by Hosoya and collaborators \cite{hosoya}. 

Let us consider a black hole in thermal equilibrium with 
a heat bath at the Hawking temperature $T_{bh}$.  
This is the Hartle-Hawking state \cite{hh}, and can be
rendered stable by placing the black hole in a cavity 
whose dimensions are very much larger than the radius of the black 
hole event horizon, thereby forming  a closed system.
We consider a small quantum system outside the black hole,
having Hamiltonian $H$ and initially in a quantum
state described by the density matrix $\rho_i$.   
We then suppose that the small quantum system 
comes into thermal equilibrium, so that its final state is the
thermal state $\rho_f=Z^{-1} e^{-\beta_{bh}  H}$, where   
$Z=tr [ e^{-\beta_{bh} H}]$ and $\beta_{bh} = \frac{1}{T_{bh}}$ 
(we have set $k_{B}$, the Boltzmann constant, equal to unity).
If the cavity is sufficiently large and the quantum system small,
we may suppose that the black hole temperature is not affected 
by this process, so that we are concerned only with quasistationary
black holes.  
Then the change of entropy outside the black hole 
is simply the entropy difference between the initial and final  
states which is  
\be 
\Delta S_{out} = S(\rho_f)-S(\rho_i)= 
tr[-\rho_f \log \rho_f + \rho_i \log \rho_i ] .
\ee 
In other words, the amount of entropy $\Delta S_{out}$ has been 
dropped into the black hole. 

In order to evaluate the change 
of black hole entropy,  we first calculate the change
in energy outside the black hole.
Then, by conservation of energy, this will be minus the
change in energy of the black hole.
Then, using the usual first law of thermodynamics,
dividing by the black hole temperature (and the Boltzmann constant), 
the change in black hole entropy is given by
\be 
\Delta S_{bh} = -\beta_{bh} tr[ H(\rho_f - \rho_i )] .  
\ee
Since $\rho_f$ is a thermal state, 
$H=-\beta _{bh}^{-1} \log (Z\rho_f ) $, which gives
\bea
\Delta S_{bh} & = &
tr[\rho _{f} \log (Z\rho _{f})] -tr [\rho _{i} \log (Z\rho _{f}) ]
\nonumber \\
& = & -tr [ (\rho _{i} - \rho _{f} ) \log \rho _{f} ]
+[ tr (\rho _{f}) - tr (\rho _{i})] \log Z
\nonumber \\
&=& -tr [ (\rho_i -\rho_f ) \log \rho_f ] .
\label{Sbh}
\eea 
The final line follows by conservation of probability.
Note that this does not assume that the states evolve unitarily.

Therefore the total change in entropy
can be written as follows 
\be 
\Delta S_{out} + \Delta S_{bh} =
tr [ \rho_i \log \rho_i - \rho_i \log \rho_f ]  .
\label{Sout}
\ee 
At this stage it is important to note that, in common with other
proofs of the generalized second law, we have had to use the first
law.
Here we have used the ordinary first law of thermodynamics, 
although this is directly analogous to the first law of
black hole mechanics for quasistationary black holes, and 
gives the same result.

We should also emphasise at this stage that the process we are
considering here is different from the usual gedanken
experiment of Bekenstein \cite{bekenstein1} in which a system
containing entropy is dropped down the black hole horizon.
Here we consider instead a system which comes into thermal 
equilibrium with the radiation outside the black hole
event horizon.
This system will be pertinent to our subsequent 
consideration of information erasure and Landauer's principle.

The quantity (\ref{Sout}) is known as quantum relative entropy 
 which is defined as  
$S(\sigma ||\rho ) =tr [\sigma \log \sigma - \sigma \log \rho ]$.  
 Quantum relative entropy $S(\sigma ||\rho)$  
has been shown \cite{qre} to be always non-negative and 
is zero if and only if $\sigma = \rho$.
Therefore (\ref{Sout}) is non-negative and 
this proves the generalized second law.

Quantum relative entropy has been shown to have 
various applications in quantum 
information theory (see \cite{schumacher}, for example) 
including entanglement quantification.
We now give a simple example to illustrate this concept.  
The unit of quantum 
information is called a quantum bit or qubit.  
A qubit is a superposition of $|0\RR$ and $|1\RR$, 
an orthonormal basis in a two-dimensional Hilbert space.  
For example, a spin-$\frac{1}{2}$ state can be considered 
as a qubit where $|0\RR$ and $|1\RR$ are spin up and down states.  
An entanglement of two qubits in subsystems $A$ and $B$  
can be written as $|\psi\RR_{AB} = a|00\RR_{AB} + b|11\RR_{AB}$  
where $a^2+b^2=1$ (if we assume $a,b$ are real). 
Then the von Neumann entropy of the reduced density matrix of 
$|\psi\RR_{AB}$, which is $-a^2 \log a^2 - b^2 \log b^2$,
yields a good measure of the entanglement.  
However it is not easy  to determine the  
amount of entanglement for mixed states with von Neumann entropy.  
Relative entropy, $S(\sigma ||\rho)$, has been shown \cite{vedral1} 
to be useful in quantifying entanglement 
for both pure and mixed states.
For pure states, relative entropy 
would reduce to the von Neumann entropy.  
Let us consider, as an example, a $|\psi\RR_{AB}$ 
which in density matrix 
form is written as    
\be 
\sigma_{AB} = \left( \begin{array}{cc} a^2 & ab 
\\ ab & b^2 \end{array} \right) .
\label{sigma}
\ee 
In order to give a correct measure of entanglement, 
$\rho_{AB}$ in $S(\sigma_{AB} ||\rho_{AB})$ satisfies the
following conditions:
 (1) it  is disentangled (i.e. 
$\rho_{AB} = \sum_i p_i \rho_A^i \otimes \rho_B^i $)  and 
 (2)
$S(\sigma_{AB} || \rho_{AB})$ is minimal.  
For pure states, a $\rho_{AB}$ satisfying both these conditions can 
always be  found as $\sigma_{AB}$ with off-diagonal 
terms set to zero, i.e.  
\be 
\rho_{AB} = \left( 
\begin{array}{cc} 
a^2 & 0 \\ 0 & b^2 
\end{array} \right) .  
\label{rho}
\ee 
With $\sigma_{AB}$ and $\rho_{AB}$ in (\ref{sigma}) and (\ref{rho}),  
we could calculate 
$tr [\sigma_{AB}\log \sigma_{AB}-\sigma_{AB} \log \rho_{AB}]$ where   
$tr [\sigma_{AB} \log \sigma_{AB}]$ is zero since $\sigma_{AB}$ 
is a pure state. 
The second term, 
$tr [-\sigma_{AB} \log \rho_{AB}]$ yields   
$-a^2 \log a^2 - b^2 \log b^2$ which is same as the von 
Neumann entropy of $|\psi\RR_{AB}$.

We shall now relate our black hole process and the
generalized second law to
Landauer's principle of information erasure.
First, we briefly review some of the key ideas, and a mechanism
for the erasure of information. 
Landauer's principle of information erasure 
(that the erasure of a certain amount of
information produces at least the equivalent amount of entropy) 
has been used to 
explain some of the fundamental aspects in quantum information theory.  
The entanglement shared by two parties can be manipulated 
into another state by local operation and classical  
communication (LOCC).  
However it is known that 
local operation cannot increase the entanglement shared by  
two separated parties.  
For example, the conversion from  
$a|00\RR_{AB} + b|11\RR_{AB}$, where $a,b \neq \frac{1}{\sqrt 2}$,  
to $\frac{1}{\sqrt 2} (|00\RR_{AB} + |11\RR_{AB}$) 
(known as entanglement purification)  
cannot be done with probability 1 by LOCC because the 
entanglement has increased  from $-a^2 \log a^2-b^2 \log b^2$ 
to $\log 2$.   
Vedral \cite{vedral2} has shown that Landauer's 
principle yields an upper bound  for entanglement purification, 
linking no local increase of entanglement to the second law of 
thermodynamics. 
Another fundamental idea in quantum information 
theory is the Holevo bound \cite{holevo} which limits the   
amount of classical information encoded in quantum mixed states 
that can be recovered.  
Plenio showed \cite{plenio} how this Holevo 
bound may be illustrated using Landauer's principle. 
In the following, we show how information 
erasure may be realized physically,
following Lubkin's method \cite{lubkin}, as  
presented in \cite{vedral2,plenio}.
We refer the reader to \cite{plenio} for details of the method.
  
Let us consider a quantum state of a system $S$ 
\be 
|\psi\RR_S = \sum_i \sqrt{\lambda_i}|a_i\RR , 
\ee 
and an apparatus $M$, initially in some pure state.  
In order to make a measurement, $M$ interacts with the state  
$|\psi\RR_S$ and entangles itself as 
\be 
\sum_i\sqrt{\lambda_i}|a_i\RR_S |m_i\RR_M  .
\label{entangled}
\ee 
The state of the the apparatus $M$  
can be obtained by tracing over the system $S$ in (\ref{entangled}),  
which yields  
\be
\rho=\sum_i \lambda_i |m_i\RR\LL m_i|,
\label{mixed}
\ee 
i.e. with probability $\lambda_i$ the apparatus is 
in state $|m_i\RR$. 
After the measurement, the apparatus will therefore
be in one of these pure states, with the associated 
probability.
The general way to erase the information of apparatus is 
to put the apparatus into a thermal reservoir.
The apparatus reaches thermal equilibrium with the reservoir 
and we then bring in another pure state to perform the next measurement.  
Then the erasure entropy has two parts: one is the entropy 
change of apparatus due to its change of state from one
of the pure states in (\ref{mixed})
to a state which is in thermal equilibrium,
and the other is the entropy change  
of the reservoir due to the apparatus.
  
In order to make this information erasure process
applicable to black holes,  
we choose a reservoir which is at the
black hole temperature, $T_{bh}$.  
Then, as shown in Figure 1, the apparatus $M$ in state $\rho$ is 
thrown into the reservoir with temperature $T_{bh}$.  
\begin{figure}
\begin{center}
\setlength{\unitlength}{1mm}  
\begin{picture}(60,40)    
\put(0,0){\framebox(60,40)} 
\put(30,10){\oval(25,15)} 
\put(45,25){$T_{\rm bh}$} 
\put(2,35){Reservoir} 
\put(21,11){Apparatus} 
\put(28,5){$M$}  
\end{picture} 
\caption{ 
The apparatus $M$ is thrown into a reservoir 
with black hole temperature $T_{bh}$ and then 
reaches thermal equilibrium. }
\end{center}
\end{figure}
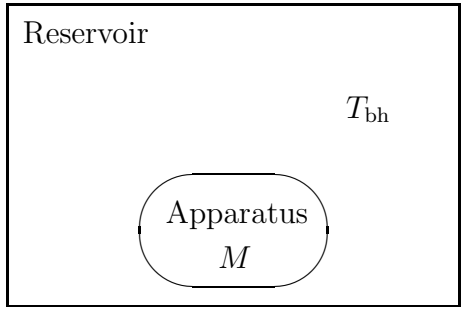
After the apparatus is thrown into the reservoir, it reaches 
thermal equilibrium and the state becomes  
\be 
\omega = Z^{-1} e^{-\beta _{bh}H} 
\ee 
where $\beta _{bh}= \frac{1}{T_{bh}}$, with $H$ the  
Hamiltonian of the apparatus, and the partition 
function is $Z= tr[ e^{-\beta _{bh}H}]$.
We imagine the process divided into two parts, firstly the
entropy of the apparatus is reduced to zero (destroying the
information in the apparatus), and then increased
by the erasure process from zero to its final value.  
This is equivalent to the process described in
\cite{plenio}, where the apparatus is in a pure
state before the erasure takes place.
The erasure entropy is given by the sum of the 
entropy changes of the apparatus and the reservoir.
The entropy change of the apparatus is simply \cite{plenio} 
(since we have already reduced its entropy to zero) 
\be 
\Delta S_{app} = -tr[\omega \log \omega] .
\label{Sapp}
\ee 
The entropy change of the reservoir can be obtained by
a method similar to that used previously for the black hole,
again assuming that the apparatus is much smaller than the reservoir
so that the temperature is not altered. 
We evaluate the heat change of the reservoir and then divide by the   
black hole temperature (and Boltzmann constant), to give  
\bea 
\Delta S_{res} &=& -\beta \{ tr[\omega H]-tr[\rho H] \}  
\nonumber 
\\      
&=& tr[ \omega \log (\omega Z)]-tr[\rho \log (\omega Z)] 
\nonumber 
\\      
&=& tr[(\omega - \rho ) \log \omega ]  ,
\label{Sres}
\eea 
where we have again used conservation of probability.
The entropy of erasure is then   
\bea 
\Delta S_{era}& =& \Delta S_{app} + \Delta S_{res}  
\\                 
&=& -tr [\rho \log \omega ] .  
\eea 
Therefore if we consider the entropy of the lost information as 
$\Delta S_{inf} = 0 - (-tr[\rho \log \rho])$, then  
\be 
\Delta S_{inf} + \Delta S_{era} = 
tr[ \rho \log \rho - \rho \log \omega] . 
\label{Sinf}
\ee 
With the identification of $\rho$ and $\omega$ as $\rho_i$ and 
$\rho_f$ in the black hole case, respectively,  
(\ref{Sinf}) yields the same result as in (\ref{Sout}). 

Application of Landauer's principle then tells us that
(\ref{Sinf}) must be positive, confirming our
proof of the generalized second law.  
However note that $\Delta S_{era}$ is not equal to $\Delta S_{bh}$,
contrary to intuition.
The reason for this lies in the details of the erasure process
\cite{plenio}.
As described above, the apparatus is in a pure state
before the erasure process takes place, so
that the change in entropy of the apparatus due
to the erasure process (\ref{Sapp}) involves
only the change between this pure state and the thermal
state $\omega $, rather than between the thermal state
and the initial state $\rho $.

Our two approaches in this paper are therefore complementary.
In the first method, we change the entropy of the system outside
the black hole, and there is a corresponding change in the entropy of the
black hole.
In the second scenario, we work out how much information we are losing
in terms of destroying the initial state, and then calculate
the amount of entropy required in order to erase this amount of 
information.
Note that in the black hole situation, the information is effectively
destroyed because the final state of the quantum system is the same thermal
state as the surrounding radiation.
We also emphasise that in the first scenario all the 
entropy lost goes down the black hole event horizon,
and the entropy of the thermal radiation surrounding the black hole
does not change.

In conclusion, in this paper
we have considered the generalized second law from the point
of view of quantum information theory, especially information erasure
and Landauer's principle.
We have considered a quantum system outside a black hole, which
comes into thermal equilibrium with the radiation
surrounding the event horizon.
For this situation, we have been able to give 
a simple proof of the generalized second law
of black hole thermodynamics by appealing to known
results on quantum relative entropy. 
This result is confirmed by an analysis of
the corresponding information erasure process 
using Landauer's principle.
This illustrates the power of quantum information theoretic
ideas when applied to black hole processes. 
\vspace{1cm}
\newline
{\bf Acknowledgment} $\;$ DS is grateful to Sougato Bose, Lucien Hardy, 
Vlatko Vedral and Ernesto Galv\~{a}o for valuable discussions. 
    
\end{document}